\documentclass[multphys,vecphys]{svmult}

\usepackage{makeidx}     
\usepackage{graphicx}    
\usepackage{multicol}    

\begin{document}

\title*{OH Absorption in the Line of Sight Towards Sagittarius B2}
\author{E. T. Polehampton\inst{1}\and J.-P. Baluteau\inst{2}\and  B. M. Swinyard\inst{3}}
\institute{Max-Planck-Institut f\"{u}r Radioastronomie, Bonn, Germany (\texttt{epoleham@mpifr-bonn.mpg.de})
\and Laboratoire d'Astrophysique de Marseille, France (\texttt{Jean-Paul.Baluteau@oamp.fr}) 
\and Rutherford Appleton Laboratory, Oxfordshire, UK (\texttt{B.M.Swinyard@rl.ac.uk})}

\maketitle

\section{Abstract}
The giant molecular cloud complex, Sagittarius B2 (Sgr B2), was observed with the ISO Long Wavelength Spectrometer over the range 47--197~$\umu$m. The ground state rotational transitions of OH lie in the survey range and we have analysed results from the $^{16}$OH, $^{17}$OH and $^{18}$OH isotopomers. Absorption in these lines due to clouds along the line of sight towards Sgr B2 has allowed us to compare the isotopomer abundance at different galactocentric radii. This is an excellent test of previous results for the isotope ratios which were measured at radio wavelengths and generally required a double ratio with $^{12}$C/$^{13}$C.

\section{Introduction}

Sgr B2 was observed over the wavelength range 47--197~$\umu$m in a full and unbiased spectral survey using the Infrared Space Observatory Long Wavelength Spectrometer (LWS). A spectral resolution of 30--40~km~s$^{-1}$ was achieved across this range using the instrument in its Fabry-P\'{e}rot mode, L03. Lines observed in the survey include rotational transitions of molecules (e.g. OH, H$_{2}$O, H$_{3}$O$^{+}$, NH, NH$_{2}$, NH$_{3}$, CH) as well as atomic fine structure lines (e.g. [OI], [OIII], [CII], [NII], [NIII]). All of the low lying rotational transitions of OH were covered by the survey.

Sgr~B2 is a giant molecular cloud complex, located close to the Galactic Centre. It emits a strong far-infrared (FIR) continuum spectrum with a peak near 80~$\umu$m \cite{goicoechea_c}. This is a perfect background against which to observe line of sight absorption features. 

\begin{figure}[!t]
\centering
\includegraphics[height=6cm]{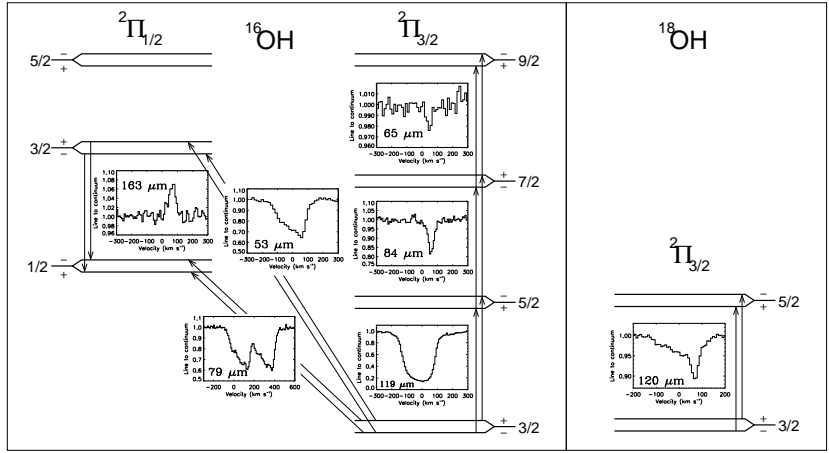}
\caption{LWS Fabry-P\'{e}rot observations of $^{16}$OH (left) and $^{18}$OH (right) shown with their rotational levels. The $J$ value of each level is indicated. The two resolved $\varLambda$-components have been averaged together for each transition except at 79~$\umu$m.}
\label{obs_lev}       
\end{figure}

We detected several $^{16}$OH transitions as well as one $^{18}$OH transition. Each line shows two resolved $\varLambda$-doublet components. Figure~\ref{obs_lev} shows the detected rotational transitions. Transitions from the lowest level show absorption due to the whole line of sight in the range $-107$~km~s$^{-1}$ to $+30$~km~s$^{-1}$. This is due to galactic spiral arm clouds located between the Sun and Galactic Centre \cite{greaves94}. Higher transitions are only seen at the velocity of Sgr B2 itself ($\sim $+65~km~s$^{-1}$; \cite{martin-pintado}) and these have been modelled by \cite{goicoechea_b}.

\section{Modelling of Ground State Line Shapes}

We modelled the observed shape of the $^{16}$OH and $^{18}$OH ground state lines at 79~$\umu$m, 53~$\umu$m and 120~$\umu$m accounting for 10 line of sight features. We used HI 21cm absorption measurements from \cite{garwood} as the basis for the model to give velocities and line widths for each feature. In the model we adjust optical depths and $^{16}$OH/$^{18}$OH ratios to simultaneously fit the three lines. 

The highest column density is associated with Sgr B2 at $+67$~km~s$^{-{1}}$: N($^{16}$OH)$\sim3\times10^{16}$~cm$^{-2}$ and N($^{18}$OH)$\sim1\times10^{14}$~cm$^{-2}$.

Two HI clouds show very little OH ($-52$~km~s$^{-1}$ \& $+53$~km~s$^{-1}$) but in the remaining line of sight clouds the average column densities are: N($^{16}$OH)$\sim5\times10^{15}$~cm$^{-2}$ and N($^{18}$OH)$\sim1\times10^{13}$~cm$^{-2}$.

\section{$^{16}$O/$^{18}$O Ratios}

Measurements of isotopic ratios through the Galaxy probe its chemical evolution as they are determined by stellar processing and outflow mechanisms. $^{16}$O is a primary product of stellar nucleosynthesis, produced directly from the primordial elements H and He, whereas $^{17}$O and $^{18}$O are secondary products which require heavier elements from previous nuclear burning for their production (see e.g. \cite{wilsonb}). Due to the higher stellar processing rates towards the Galactic Centre, the $^{16}$O/$^{18}$O ratio is expected to be lower there than in the Solar neighbourhood.

The rotational transitions of OH are a particularly good tracer for the oxygen isotope ratio $^{16}$O/$^{18}$O because there are no efficient chemical reactions causing isotopic fractionation that would change $^{16}$OH/$^{18}$OH from the true value \cite{langer}. Also, the transitions in both $^{16}$OH and $^{18}$OH are optically thin for line of sight clouds and in these clouds the ground state lines are a good measure of the total column density (absorption from higher levels is not observed).

\begin{figure}[!t]
\centering
\includegraphics[height=6cm]{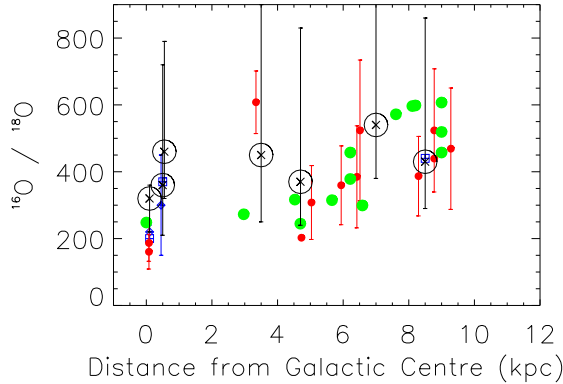}
\caption{Preliminary $^{16}$OH/$^{18}$OH ratios for 7 line of sight features plotted against distance from Galactic Centre (large circles with cross). Values from the literature are plotted for comparison. Red filled circles from \cite{kahane}, green filled circles from \cite{wilson}, blue squares with cross from \cite{bujarrabal} and blue diamonds from \cite{williams}.}
\label{results}       
\end{figure}

Most previous measurements of this ratio have been determined from radio lines using H$_{2}^{13}$C$^{16}$O/H$_{2}^{12}$C$^{18}$O \cite{wilson}. This a double ratio with $^{12}$C/$^{13}$C to avoid optically thick transitions. The FIR lines of OH provide an important test of these values with a different molecule in a different wavelength regime and without reference to $^{12}$C/$^{13}$C (which is subject to fractionation effects). Preliminary results from the fit are given in Fig.~\ref{results}, showing good consistency with previous values. A first estimate of the modelling errors are given.

\section{$^{17}$OH}

\begin{figure}[!t]
\centering
\includegraphics[height=5cm]{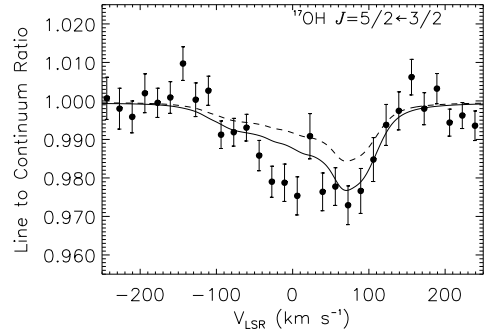}
\caption{$^{17}$OH absorption line data with model derived from the observed $^{18}$OH absorption and $^{18}$O/$^{17}$O ratio of 3.5 (solid line). The predicted shape using the Solar System $^{18}$O/$^{17}$O ratio of 5.2 is also shown for comparison (dashed line).}
\label{17oh}      
\end{figure}

$^{17}$OH was also detected in the survey via its lowest energy rotational transition, the first detection of this line in the ISM \cite{polehampton}. This transition is interesting because the $^{17}$O nucleus has non-zero spin and so causes extra splitting into hyperfine components spread over $\sim$80~km~s$^{-1}$, modifying the line shape. The observed absorption is greater than predicted using the Solar System $^{18}$O/$^{17}$O ratio (5.2) and at the velocity of Sgr B2 it is consistent with previous measurements of $^{18}$O/$^{17}$O through the Galaxy ($\sim$3.5 \cite{penzias}) - see Fig.~\ref{17oh}.

\end{document}